\documentclass{svproc}
\usepackage{url}

\begin{document}
\mainmatter              % start of a contribution
\title{Four-body Faddeev-type equations for $\bar{K}NNN$ quasi-bound state
calculations}
\titlerunning{Four-body Faddeev-type equations}  % abbreviated title (for running head)
\author{Nina Shevchenko\inst{1}}
\authorrunning{Nina Shevchenko} % abbreviated author list (for running head)
\tocauthor{Nina Shevchenko}
\institute{Nuclear Physics Institute, \v{R}e\v{z} 250 68, Czech Republic\\
\email{shevchenko@ujf.cas.cz}
%WWW home page:
%\texttt{http://users/\homedir iekeland/web/welcome.html}
}

\maketitle              % typeset the title of the contribution

\begin{abstract}
The paper is devoted to the $\bar{K}NNN$ system, which is an exotic system consisting
of an antikaon and three nucleons. Four-body Faddeev-type AGS equations, which are being
used for evaluation of the possible quasi-bound state in the system are described.
% We would like to encourage you to list your keywords within
% the abstract section using the \keywords{...} command.
\keywords{few-body physics, antikaon-nucleon systems, four-body Faddeev equations}
\end{abstract}
\section{Introduction}

The attractive nature of $\bar{K}N$ interaction has stimulated theoretical and experimental
searches for bound states of $K^-$ with different number of nucleons. The interest
in few-body antikaonic systems was stimulated by calculations, which predicted deep and relatively
narrow quasi-bound $K^-$ -nuclear states. Many theoretical calculations devoted to the lightest
possible system $\bar{K}NN$ have been performed since then using different methods, see e.g.
a review \cite{review}. All of them agree that a quasi-bound state in the $K^- pp$ system exists,
but they yield quite diverse binding energies and widths.

Some theoretical works were devoted to the question of the quasi-bound state
in the four-body $\bar{K}NNN$ system, but more accurate calculations within
Faddeev-type equations are needed. Indeed, only these dynamically exact equations
in momentum representation can treat energy dependent $\bar{K}N$ potentials, necessary
for the this system, exactly.

The paper contains description of the four-body Faddeev-type AGS equations \cite{4AGS},
written down for the $\bar{K}NNN$ system. We will solve the equations using our programs
written for the three-body AGS calculations of the $\bar{K}NN$ system, described in \cite{review},
and our two-body potentials constructed for them.

\section{Four-body Faddeev-type AGS equations}

The four-body AGS equations contain three-body transition operatos, obtained from
the three-body AGS equations. The three-body Faddeev-type equations in AGS
form~\cite{AGS} written for separable potentials
 $V_{\alpha} = \lambda_{\alpha} |g_{\alpha} \rangle \langle g_{\alpha}|$
have a form
\begin{equation}
\label{3AGSsep}
 X_{\alpha \beta}(z) = Z_{\alpha \beta}(z) + 
 \sum_{\gamma=1}^3 Z_{\alpha \gamma}(z) \tau_{\gamma}(z) X_{\gamma \beta}(z)
\end{equation}
with transition $X_{\alpha \beta}$ and kernel $Z_{\alpha \beta}$ operators
\begin{eqnarray}
\label{X3b}
 X_{\alpha \beta}(z) &=&
  \langle g_{\alpha} | G_0(z) U_{\alpha \beta}(z) G_0(z) | g_{\beta} \rangle, \\
\label{Z3b}
 Z_{\alpha \beta}(z) &=& (1-\delta_{\alpha \beta})
   \langle g_{\alpha} | G_0(z) | g_{\beta} \rangle.
\end{eqnarray}
Here $U_{\alpha \beta}(z)$ is the three-body transition operator, which describes process
$\beta + (\alpha \gamma) \to \alpha + (\beta \gamma)$, while $G_0(z)$ is the three-body
Green function. Faddeev partition indices $\alpha, \beta = 1,2,3$ simultaneously define
a particle ($\alpha$) and the remained pair ($\beta \gamma$).
The operator $\tau_{\alpha}(z)$ in Eq.(\ref{3AGSsep}) is an energy-dependent part of a separable
two-body $T$-matrix
$
T_{\alpha}(z) = |g_{\alpha} \rangle \tau_{\alpha}(z) \langle g_{\alpha}|,
$
describing interaction in the ($\beta \gamma$) pair; $| g_{\alpha} \rangle$ is a form-factor

%------------------------------------------------------------------------------
The four-body Faddeev-type AGS equations~\cite{4AGS}, written for separable potentials, have a form
\begin{eqnarray}
\label{4AGSsepV}
 \bar{U}^{\sigma \rho}_{\alpha \beta}(z) &=& (1-\delta_{\sigma \rho}) 
  (\bar{G_0}^{-1})_{\alpha \beta}(z) + 
 \sum_{\tau,\gamma,\delta} (1-\delta_{\sigma \tau}) \bar{T}^{\tau}_{\alpha \gamma}(z)
  (\bar{G_0})_{\gamma \delta}(z) \bar{U}^{\tau \rho}_{\delta \beta}(z), \\
\label{barU}
 &{}& \bar{U}^{\sigma \rho}_{\alpha \beta}(z) = 
  \langle g_{\alpha} | G_0(z) U^{\sigma \rho}_{\alpha \beta}(z) G_0(z) | g_{\beta} \rangle, \\
\label{barT}
  &{}& \bar{T}^{\tau}_{\alpha \beta}(z) = 
  \langle g_{\alpha} | G_0(z) U^{\tau}_{\alpha \beta}(z) G_0(z) | g_{\beta} \rangle, 
\;
 (\bar{G_0})_{\alpha \beta}(z) = \delta_{\alpha \beta} \tau_{\alpha}(z).
\end{eqnarray}
The operators $\bar{U}^{\sigma \rho}_{\alpha \beta}$ and $\bar{T}^{\tau}_{\alpha \beta}$
contain four-body $U_{\alpha \beta}^{\sigma \rho}(z)$ and three-body $U_{\alpha \beta}^{\tau}(z)$
transition operators, correspondingly. The last one ($U_{\alpha \beta}^{\tau}(z)$) differs
from the three-body operator in Eq.(\ref{X3b}) by additional upper index $\tau$ ($\sigma$, $\rho$),
which defines a three-body subsystem of the four-body system. The free Green function $G_0(z)$
now acts in four-body space. If, in addition, the ''effective three-body potentials''
$\bar{T}^{\tau}_{\alpha \beta}(z)$ in Eq.~(\ref{4AGSsepV}) are presented in a separable form:
$
\bar{T}^{\tau}_{\alpha \beta}(z) =  | \bar{g}^{\tau}_{\alpha} \rangle 
  \bar{\tau}^{\tau}_{\alpha \beta}(z)   \langle \bar{g}^{\tau}_{\alpha} |,
$
the four-body equations can be written as \cite{4AGSsep}
\begin{equation}
\label{4AGSsepVT}
 \bar{X}^{\sigma \rho}_{\alpha \beta}(z) = \bar{Z}^{\sigma \rho}_{\alpha \beta}(z) + 
 \sum_{\tau,\gamma,\delta} \bar{Z}^{\sigma \tau}_{\alpha \gamma}(z) \bar{\tau}^{\tau}_{\gamma \delta}(z)
   \bar{X}^{\tau \rho}_{\delta \beta}(z)
\end{equation}
with new transition $\bar{X}^{\sigma \rho}$ and kernel $\bar{Z}^{\sigma \rho}$ operators
defined by
\begin{eqnarray}
 \bar{X}^{\sigma \rho}_{\alpha \beta}(z) &=&
  \langle \bar{g}^{\sigma}_{\alpha} | \bar{G_0}(z)_{\alpha \alpha} \bar{U}^{\sigma \rho}_{\alpha \beta}(z)
    \bar{G_0}(z)_{\beta \beta} | \bar{g}^{\rho}_{\beta} \rangle, \\
 \bar{Z}^{\sigma \rho}_{\alpha \beta}(z) &=& (1-\delta_{\sigma \rho})
   \langle \bar{g}^{\sigma}_{\alpha} | \bar{G_0}(z)_{\alpha \beta} | \bar{g}^{\rho}_{\beta} \rangle.
\end{eqnarray}
The separabelization of the ''effective three-body potentials'' $\bar{T}^{\tau}_{\alpha \beta}(z)$
can be performed using e.g. the Hilbert-Schmidt expansion of the three-body AGS equations with separable
potentials Eq.(\ref{3AGSsep}).

\section{Four-body equations for the $\bar{K}NNN$ system}

There are two types of partitions for a four-body system: $3+1$ and $2+2$. For
the $\bar{K}NNN$ system they are: $|\bar{K} + (NNN) \rangle$, $|N + (\bar{K}NN) \rangle$
and $|(\bar{K}N) + (NN) \rangle$. At the begin we considered all three nucleons
as different particles, so we started by writing down the four-body system of equations
Eq.(\ref{4AGSsepVT}) for the following $18$ channels $\sigma_{\alpha}$
(with $\alpha =$ $NN$ or $\bar{K}N$):
\begin{eqnarray}
\nonumber
&{}&  1_{NN}  :  |\bar{K} + (N_1 + N_2 N_3) \rangle, |\bar{K} + (N_2 + N_3 N_1) \rangle,
                      |\bar{K} + (N_3 + N_1 N_2) \rangle,          \\
\nonumber
&{}&  2_{NN}:  |N_1 + (\bar{K} + N_2 N_3) \rangle, |N_2 + (\bar{K} + N_3 N_1) \rangle,
                      |N_3 + (\bar{K} + N_1 N_2) \rangle, \\
\label{channels}                      
&{}&  2_{\bar{K}N}: |N_1 + (N_2 + \bar{K} N_3) \rangle, |N_2 + (N_3 + \bar{K} N_1) \rangle,
                      |N_3 + (N_1 + \bar{K} N_2) \rangle,\\
\nonumber                      
&{}&  \quad \quad \quad \! |N_1 + (N_3 + \bar{K} N_2) \rangle, |N_2 + (N_1 + \bar{K} N_3) \rangle,
                      |N_3 + (N_2 + \bar{K} N_1) \rangle,\\
\nonumber
&{}&  3_{NN}:  |(N_2 N_3) + (\bar{K} + N_1) \rangle, |(N_3 N_1) + (\bar{K} + N_2) \rangle,
                   |(N_1 N_2) + (\bar{K} + N_3) \rangle, \\
\nonumber                  
&{}&  3_{\bar{K}N}: |(\bar{K} N_1) + (N_2 + N_3) \rangle, |(\bar{K} N_2) + (N_3 + N_1) \rangle,
                      |(\bar{K} N_3) + (N_1 + N_2) \rangle
\end{eqnarray}
After this the operators and equations were antisymmetrized, and the system of operator equations
was written in a form:
\begin{equation}
\label{4AGSfinal}
 \hat{X} = \hat{Z} \, \hat{\tau} \, \hat{X},
\end{equation}
were $\hat{Z}$ and $\hat{\tau}$ are the $5 \times 5$ matrices containing the kernel operators
$\bar{Z}^{\sigma \rho}_{\alpha}$ and $\bar{\tau}^{\rho}_{\alpha \beta}$, correspondingly. 
Since the initial state is assumed to be fixed, only one column of the $5 \times 5$ matrix $\hat{X}$,
containing transition operators $\bar{X}^{\sigma \rho}_{\alpha \beta}$, is necessary:
\begin{equation}
\label{XZ5x5}
 \bar{X}^{\rho}_{\alpha} = \left(
 \begin{tabular}{c}
  $\bar{X}^{1}_{NN}$ \\
  $\bar{X}^{2}_{NN}$ \\
  $\bar{X}^{2}_{\bar{K}N}$ \\
  $\bar{X}^{3}_{NN}$ \\  
  $\bar{X}^{3}_{\bar{K}N}$ \\  
 \end{tabular}
 \right),
%=========================================
\qquad \qquad
%=========================================
 \bar{Z}^{\sigma \rho}_{\alpha} = \left(
 \begin{tabular}{ccccc}
  0 & $\bar{Z}^{12}_{NN}$ & 0 & $\bar{Z}^{13}_{NN}$ & 0 \\
  $\bar{Z}^{21}_{NN}$ & 0 & 0 & $\bar{Z}^{23}_{NN}$ & 0 \\
  0 & 0 & $\bar{Z}^{22}_{\bar{K}N}$ & 0 & $\bar{Z}^{23}_{\bar{K}N}$ \\
  $\bar{Z}^{31}_{NN}$ & $\bar{Z}^{32}_{NN}$ & 0 & 0 & 0 \\  
  0 & 0 & $\bar{Z}^{32}_{\bar{K}N}$ & 0 & 0 \\  
 \end{tabular}
 \right),
\end{equation}
%---------------------------------------------------------------------
\begin{equation}
\label{tau5x5}
 \bar{\tau}^{\rho}_{\alpha \beta} = \left(
 \begin{tabular}{ccccc}
  $\bar{\tau}^{1}_{NN,NN}$ & 0 & 0 & 0 & 0 \\
  0 & $\bar{\tau}^{2}_{NN,NN}$ & $\bar{\tau}^{2}_{NN,\bar{K}N}$ & 0 & 0 \\
  0 & $\bar{\tau}^{2}_{\bar{K}N,NN}$ & $\bar{\tau}^{2}_{\bar{K}N,\bar{K}N}$ & 0 & 0 \\
  0 & 0 & 0 & $\bar{\tau}^{3}_{NN,NN}$ & $\bar{\tau}^{3}_{NN,\bar{K}N}$ \\  
  0 & 0 & 0 & $\bar{\tau}^{3}_{\bar{K}N,NN}$ & $\bar{\tau}^{3}_{\bar{K}N,\bar{K}N}$ \\  
 \end{tabular}
 \right).
\end{equation}

\section{Three-body subsystems and two-body input}

We are studying the $\bar{K}NNN$ system with the lowest value of the four-body isospin
$I^{(4)}=0$, which can be denoted as $K^- ppn$. Its total spin $S^{(4)}$ is equal to one half,
while the orbital momentum is zero, since all two-body interactions are chosen to be zero.
For the $\bar{K}NNN$ system with these quantum numbers the following three-body subsystems
contribute:
\begin{itemize}
 \item $\bar{K}NN$ with $I^{(3)} = 1/2, S^{(3)} = 0$ ($K^- pp$) or $S^{(3)} = 1$ ($K^- d$).
 \item $NNN$ with $I^{(3)} = 1/2, S^{(3)} = 1/2$ ($^{3}$H or $^{3}$He),
\end{itemize}
were $I^{(3)}$ and $S^{(3)}$ are the three-body isospin and spin. 

The three-body antikaon-nucleon system $\bar{K}NN$ with different quantum numbers was studied
in our previous works, see Ref.\cite{review}. In particular, quasi-bound state pole positions
in the $K^- pp$ system ($\bar{K}NN$ with $I^{(3)} = 1/2$, $S^{(3)} = 0$)
and near-threshold $K^- d$ scattering amplitudes ($\bar{K}NN$ with $I^{(3)} = 1/2$, $S^{(3)} = 1$)
were calculated (no quasi-bound states were found in the $K^- d$ system). It was
done using the three-body AGS equations with separable potentials Eq.(\ref{3AGSsep}) with three
models of the $\bar{K}N$ interaction: two phenomenological potentials having one- or two-pole
structure of the $\Lambda(1405)$ resonance and a chirally motivated model.
All three potentials describe low-energy $K^- p$ scattering and $1s$ level shift of kaonic
hydrogen with equally high accuracy. We also used a two-term separable $NN$ potential, which
reproduces Argonne v18 $NN$ phase shifts and scattering lengths. The same potentials are used
in our four-body calculations.

The programs of numerical solution of the three-body AGS equations for the $\bar{K}NN$ systems can
be used for separabelization of the ''effective three-body potentials'' (after some modifications).
However, it is far not enough since the three-body system Eq.(\ref{3AGSsep}) was solved with
the initial channel fixed by $\beta = 1$, which corresponds to the $\bar{K} + NN$ partition.
It allows to calculate the three-body transition amplitudes, which in the four-body formalism
are denoted as $\bar{T}^{2_i}_{NN,NN}$ and $\bar{T}^{2_i}_{\bar{K}N,NN}$. The other two three-body
amplitudes, necessary for the four-body calculations: $\bar{T}^{2_i}_{NN,\bar{K}N}$ and
$\bar{T}^{2_i}_{\bar{K}N,\bar{K}N}$, - were calculated additionally by solving the AGS equations
Eq.(\ref{3AGSsep}) with other two initial channels $\beta = 2,3$, corresponding to the $N + \bar{K}N$
initial state (which were properly antisymmetrized).

The three-nucleon system $NNN$ was treated by solving the system of AGS equations
Eq.(\ref{3AGSsep}) with our separable two-term $NN$ potential. The calculated binding energy was
found to be $9.95$ MeV for both ${}^3$H and ${}^3$He nuclei since Coulomb interaction was not taken
into account.

\section{Summary}

The four-body Faddeev-type calculations of the $\bar{K}NNN$ system is very complicated and
time-consuming task. Up to now the four-body AGS equations were written down for the $\bar{K}NNN$
system and antisymmetrized. All necessary calculations of the three-body (sub)systems were performed,
their separable forms were numerically evaluated. The spin-isospin re-coupling parts of the four-body
kernel functions $\bar{Z}^{\sigma \rho}_{\alpha}$ were also calculated. The work is in progress.

%\begin{thebibliography}{00}  %for 2 digits

\end{document}